\documentclass{nature}

\usepackage{color} 
\usepackage{nicefrac}
\usepackage{amsmath}
\usepackage{amssymb}
\usepackage[hypertexnames=false]{hyperref}
\usepackage{graphicx}
\usepackage[separate-uncertainty = true]{siunitx} 
\usepackage{caption}
\usepackage{sectsty} 
\usepackage{float} 

\usepackage{booktabs}

\allsectionsfont{\sffamily} 

\usepackage{braket}

\usepackage{lipsum} 

\DeclareCaptionLabelSeparator{bar}{ \textbf{$\vert$} }
\newcommand{\Spar}{$\braket{\hat{S}_{\mathrm{\|}}}$}
\newcommand{\Sper}{$\braket{\hat{S}_{\mathrm{\perp}}}$}
\newcommand{\Sz}{$\braket{\hat{S}_{z}}$}

\newcommand{\dNN}{d_\mathrm{NN}}
\newcommand{\JNN}{J_\mathrm{NN}}
\newcommand{\Dperp}{D_\mathrm{\perp}}

\newcommand{\DperpNN}{D_\mathrm{\perp, NN}}
\newcommand{\DperpNNN}{D_\mathrm{\perp, NNN}}
\newcommand{\Dpar}{D_\mathrm{\|}}
\newcommand{\Dz}{D_\mathrm{z}}

\newcommand{\Hcal}{\mathcal{H}}	
\newcommand{\muB}{\mu_\mathrm{B}}

\definecolor{stateblue}{rgb}{0.18,0.5,.75} 
\definecolor{statered}{rgb}{0.88,0.29,.29}

\newcommand {\stateblue}[1]{{\color{stateblue} #1}} 
\newcommand {\statered}[1]{{\color{statered} #1}} 

\newcommand{\didu}[0]{\nicefrac{\mathrm{d}I}{\mathrm{d}V}}

\title{Non-collinear spin states in bottom-up fabricated atomic chains}

\author{Manuel Steinbrecher$^1$, Roman Rausch$^2$, Khai Ton That$^1$, Jan Hermenau$^1$, Alexander A. Khajetoorians$^{1,3}$, Michael Potthoff$^2$, Roland Wiesendanger$^1$, and Jens Wiebe$^1$}

\begin{document}

\maketitle

\begin{affiliations}
 \item Department of Physics, Hamburg University, 20355 Hamburg, Germany
 \item I. Institute for Theoretical Physics, Hamburg University, 20355 Hamburg, Germany
 \item Institute for Molecules and Materials (IMM), Radboud University, 6525 AJ Nijmegen, The Netherlands
\end{affiliations}

\begin{abstract}
Non-collinear spin states with unique rotational sense, such as chiral spin-spirals, are recently heavily investigated because of advantages for future applications in spintronics and information technology\cite{Menzel2012,Brataas2013, Bergmann2014} and as potential hosts for Majorana Fermions\cite{Klinovaja2013, Nadj-Perge2013, Pientka2013,Schecter2016} when coupled to a superconductor. Tuning the properties of such spin states, e.g., the rotational period and sense, is a highly desirable yet difficult task\cite{Chen2013, Hrabec2014}. Here, we experimentally demonstrate the bottom-up assembly of a spin-spiral derived from a chain of Fe atoms on a Pt substrate using the magnetic tip of a scanning tunneling microscope as a tool. We show that the spin-spiral is induced by the interplay of the Heisenberg and Dzyaloshinskii-Moriya components of the Ruderman-Kittel-Kasuya-Yosida interaction\cite{Smith1976,Fert1980} between the Fe atoms. The relative strengths and signs of these two components can be adjusted by the interatomic Fe distance\cite{Khajetoorians2016}, which enables tailoring of the rotational period and sense of the spin-spiral.
\end{abstract}

Ultra-thin layers and atomic chains of transition metal atoms on high-$Z$ substrates show a variety of non-collinear spin states\cite{Bergmann2014} ranging from spin-spirals\cite{Bode2007, Ferriani2008, Menzel2012, Menzel2014, Yoshida2012, Finco2017} to skyrmions\cite{Romming2013}. In most of the cases, these non-collinear spin states are stabilized by the so-called Dzyaloshinskii-Moriya (DM) interaction $\mathbf{D}_{ij} \left( \hat{\mathbf{S}}_\text{i} \times \hat{\mathbf{S}}_\text{j}\right)$ between neighboring spins $\hat{\mathbf{S}}_\text{i}$ and $\hat{\mathbf{S}}_\text{j}$, which is relevant due to the broken inversion asymmetry at the interface of the magnetic material and the substrate. While the usual Heisenberg interaction $J_{ij} \, \hat{\mathbf{S}}_i \cdot \hat{\mathbf{S}}_j$ favors a parallel (ferromagnetic) or antiparallel (antiferromagnetic) alignment of neighboring spins, the DM interaction favors perpendicular orientations. The interplay of these two interactions can induce a spin-spiral state, with a rotational sense determined by the orientation of $\mathbf{D}_{ij}$, and with a period given by the ratio of $D_{ij}$ and $J_{ij}$, and by the magnetic anisotropy\cite{Vedmedenko2007}.

A very appealing approach to tune such a spin-spiral state is to use the tip of a scanning tunneling microscope to construct chains atom by atom with tailored magnetic anisotropy and interactions. While numerous bottom-up magnetic chains have been investigated\cite{Hirjibehedin2006,Loth2012,Khajetoorians2012,Spinelli2014,Choi2015arXiv,Toskovic2016}, these studies have focused solely on Heisenberg interactions inducing collinear order. However, the ubiquitous Ruderman-Kittel-Kasuya-Yosida (RKKY) interaction between magnetic atoms in contact to metallic substrates additionally contains a DM component leading to non-collinear order, which is typically strong for high-$Z$ materials\cite{Smith1976,Fert1980}. Indeed, it has been shown recently that the DM component of the RKKY interaction between Fe atoms on Pt(111) is of comparable strength as the Heisenberg part\cite{Khajetoorians2016}. Most notably, due to the oscillatory nature of the RKKY coupling, the two components are adjustable in their relative strengths and signs by changing the distance between the Fe atoms\cite{Khajetoorians2016}. The use of similar material combinations of dilute transition metal chains on high-$Z$ substrates therefore promises the stabilization of adjustable spin-spiral states in bottom-up fabricated chains.

\begin{figure}[H]
 	\centering
     \includegraphics[width = .8\columnwidth]{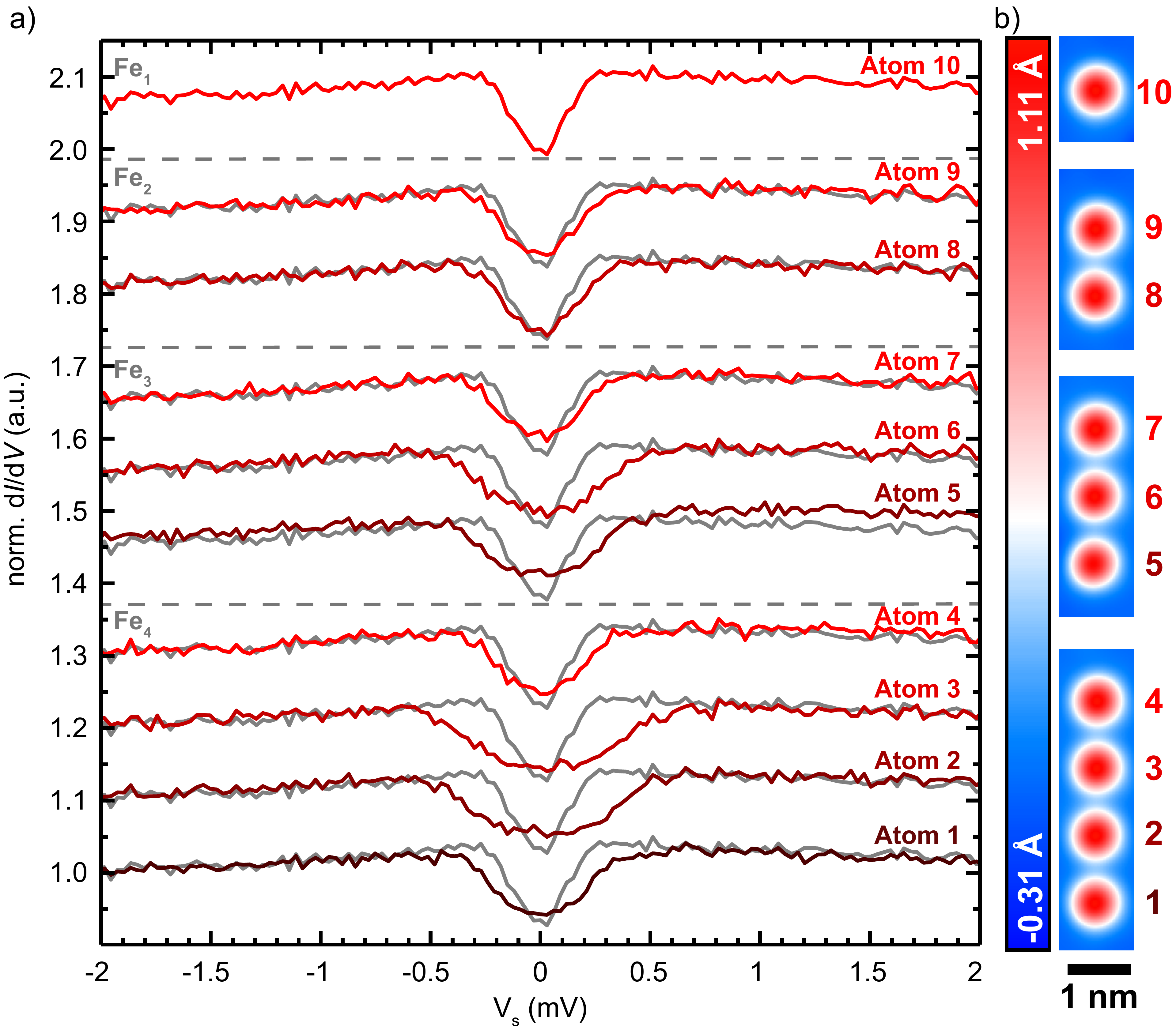}%
      \caption{\label{fig1}\textbf{ISTS of bottom-up Fe$_N$ chains of different numbers \textit{\textbf{N}} of atoms.}\\\textbf{a}, ISTS spectra taken on top of an isolated Fe atom (Fe$_1$, 10) and on the Fe atoms inside Fe$_2$ (8,9), Fe$_3$ (5,6,7), and Fe$_4$ (1,2,3,4) chains with $d_{\rm NN}=4a$. For comparison, the spectrum of the isolated Fe$_1$ is shown in gray behind all other spectra. All spectra were normalized by dividing by a substrate spectrum. ($V_\mathrm{stab} = \SI{-6}{\milli \volt}$, $I_{\mathrm{stab}} = \SI{3}{\nano\ampere}$, $V_\mathrm{mod} = \SI{40}{\micro \volt}$). \textbf{b}, STM topographs of all chains investigated in (\textbf{a}) ($V_\mathrm{s} = \SI{-6}{\milli \volt}$, $I_{\mathrm{stab}} = \SI{3}{\nano\ampere}$).}
 \end{figure}

Using tip-induced atom manipulation we built Fe$_N$ chains on a Pt(111) surface with different numbers $N$ of Fe atoms and with different nearest neighbor (NN) distances $d_{\rm NN}$ (see Methods for experimental details). \autoref{fig1}b shows an isolated Fe atom (Fe$_1$) and assembled Fe$_N$ chains with $N = 2,3,4$ and $d_{\rm NN}=4a$ with the shortest distance of equivalent hollow adsorption sites of $a=\SI{2.78}{\angstrom}$. All chains investigated in this work were built from Fe atoms sitting on the hexagonal close-packed (hcp) adsorption site, as identified by the characteristic spin-excitation at $\SI{0.19}{\milli\electronvolt}$ in inelastic scanning tunneling spectroscopy (ISTS) taken on Fe$_1$ (\autoref{fig1}a). As shown in Ref.\citenum{Khajetoorians2013}, they exhibit a relatively weak easy-plane magnetic anisotropy, which only slightly favors an orientation of the Fe spin in the surface plane. From an investigation of pairs of an Fe-hydrogen complex and a Fe atom with different distances\cite{Khajetoorians2016}, we expect that the chosen $d_{\rm NN}$ results in an antiferromagnetic Heisenberg interaction of $\JNN \approx \SI{-50}{\micro\electronvolt}$ and a comparable DM coupling $\DperpNN \approx +\SI{30}{\micro\electronvolt}$ between NN Fe atoms in the chains (see Supplementary Figure 1 for the definition of the coordinates~\cite{Supplement}).

ISTS spectra taken on the Fe atoms in the Fe$_N$ chains (\autoref{fig1}a) reveal considerable shifts of the spin-excitation to higher excitation energies and a broadening of the excitation step as compared to the isolated Fe$_1$ shown in gray as a reference\cite{Khajetoorians2016}. This effect is more pronounced for the atoms with two neighbors (atoms 2, 3, 6) and we therefore tentatively assign the shifts to the mutual magnetic interactions between the atoms. Due to symmetry arguments, the DM vector has its largest component in the surface plane perpendicular to the chain axis ($\Dpar = 0$, $\Dz$ small)\cite{Khajetoorians2016}. We therefore anticipate a tendency towards the formation of a cycloidal spin-spiral state in the ($||$,$z$)-plane. Since the chain atoms are strongly coupled to the substrate conduction electrons, these states typically do not show remanence\cite{Khajetoorians2012, Menzel2012} and therefore cannot be resolved by conventional spin-resolved STM (SPSTM) in the absence of a magnetic field. We will therefore use two different methods, namely (i) an external homogeneous magnetic field $B_z$ along the surface normal ($z$) and (ii) RKKY coupling one end of the chain to a ferromagnetic island\cite{Meier2008, Khajetoorians2011}, in order to stabilize the non-collinear spin states.


First, we applied $B_z$ to the chains and performed ISTS. For the sake of simplicity, we will focus on the data of the Fe$_4$ chain shown in \autoref{fig2}b,c (ISTS and calculations of all other chains are shown in Supplementary Note 2). The spectra reveal a small shift of the spin-excitation to lower energies for field strengths up to $B_z\approx\SI{4}{\tesla}$, and then a nearly linear increase in the spin-excitation energy proportional to $B_z$, with slight differences between the spectra on the two inner and the two edge atoms of the chain. Compared to the $B_z$-dependent ISTS of the Fe$_1$, where the minimum of the excitation energy appears at $B_z\approx \SI{3}{\tesla}$\cite{Khajetoorians2013} (see Supplementary Figure 3a), this minimum is much less pronounced and shifted to larger $B_z$ on the chain atoms. The shift again manifests the significant magnetic interactions between the Fe atoms in the chain. For a further analysis of the spin state of the chain, we quantify the spectra utilizing a perturbation theory model\cite{Ternes2015} based on an effective spin Hamiltonian that considers the Zeeman energy, the easy-plane magnetic anisotropy energy, and the Heisenberg as well as DM interactions (see Methods). Astonishingly, by simply considering the atomic parameters known from single hcp Fe atoms on Pt(111)\cite{Khajetoorians2013} and using the above Heisenberg and DM interactions extracted from the pairs\cite{Khajetoorians2016} as NN interactions in the model (see the parameters in the caption of \autoref{fig2}), we can already excellently reproduce the evolution of the excitation in the ISTS data (cf. \autoref{fig2}c and d). This result further corroborates that the pairwise interaction parameters describe the chain data fairly well. However, for this particular distance, we found that the calculated spectra for the Fe$_2$ pair and the Fe$_3$ and Fe$_4$ chains fit slightly better to the experimental data when the pairwise Heisenberg coupling\cite{Khajetoorians2016} is slightly reduced to $\JNN = \SI{-25}{\micro\electronvolt}$. This value was therefore used in all calculations for the $\dNN = 4a$ chains. 
The corresponding spin-expectation values calculated from exact diagonalization (ED, see Methods), which are illustrated in \autoref{fig2}a reveal that, as long as $B_z$ is not too large, the underlying spin state is highly non-collinear. The spins of neighboring atoms are mutually canted within the ($||$,$z$)-plane as imposed by the strong $\DperpNN$ component. Note, that a pure spin-spiral state is not realized because of the homogeneous $B_z$ driving all spins into $z$-direction (see Supplementary Note 4 and Supplementary Figure 9). We will later see, how we can avoid this problem by using an RKKY coupling of the chain end to a ferromagnetic island.


Next, in order to demonstrate the adjustability of the non-collinear chain state by $d_{\rm NN}$, we investigated Fe$_N$ chains ($N = 2,3,4$) with smaller $d_{\rm NN}=3a$ (see the ISTS data and calculations of Fe$_4$ in \autoref{fig3}, all other chains are shown in Supplementary Note 3). The smaller $d_{\rm NN}$ induces a stronger antiferromagnetic Heisenberg component ($J_{\rm NN} \approx \SI{-60}{\micro\electronvolt}$) and a stronger DM component ($\DperpNN \approx \SI{-50}{\micro\electronvolt}$) in pairs\cite{Khajetoorians2016} as compared to the $4a$-case studied above. Most notably, we know from \textit{ab-initio} calculations of the same pairs, that the sign of $\DperpNN$ is reversed\cite{Khajetoorians2016}. Indeed, $B_z$ dependent ISTS of this Fe$_4$ chain behaves rather differently compared to the $4a$-chain data (cf. \autoref{fig3}c,d and \autoref{fig2}b,c): The zero field spin-excitation appears at a larger energy, and, for the two inner atoms, the shift of the spin-excitation to lower energies continues here up to at least $B_z=\SI{7}{\tesla}$. These differences can be assigned to the change in $J_{\rm NN}$ and $\DperpNN$, as supported by the perturbation theory model calculations of the spectra shown in \autoref{fig3}e, which nicely reproduce the data. Again, we simply used the NN interactions extracted from the according pair with $d=3a$, but additionally took into account the next nearest neighbor (NNN) interactions extracted from the couplings of a pair with $d=6a$\cite{Khajetoorians2016} (see the values given in the caption of \autoref{fig3}). As corroborated by ED, the underlying spin state of the chain stabilized in $B_z=\SI{3}{\tesla}$ (see \autoref{fig3}b) is again strongly non-collinear.
Moreover, the character of the two non-collinear spin states of the $3a$- and $4a$-chains is obviously different (cf. \autoref{fig2}a to \autoref{fig3}b). While the spins of the $1^{\rm st}$ and $3^{\rm rd}$ ($2^{\rm nd}$ and $4^{\rm th}$) atom in the $4a$-chain are tilted to the left (right) (cf. \autoref{fig2}a), this tilt is opposite for the $3a$-chain (cf. \autoref{fig3}b). This is a result of the sign change of $\DperpNN$ between the two cases, which will lead to a different rotational sense if a pure spin-spiral state was stabilized. We finally allude to a subtle difference between the experimental (\autoref{fig3}d) and calculated (\autoref{fig3}e) ISTS. While the conductance levels at positive and negative bias above the excitation threshold are the same in the calculations, there is a systematic difference in the experimental data. For $B_z<\SI{2}{\tesla}$ the positive-bias conductance is lower (higher) for the $1^{\rm st}$ and $3^{\rm rd}$ ($2^{\rm nd}$ and $4^{\rm th}$) atom, and this contrast is reversed at $B_z=\SI{3}{\tesla}$. Considering the alternating tilts of the spins in the $||$-direction (cf. \autoref{fig3}b), we assign this asymmetry to a non-zero $||$-component of the spin-polarization of this particular tip, which is reversing at $B_z=\SI{3}{\tesla}$. This conclusion is further supported by the alternating heights of neighboring atoms in the SPSTM image taken on this Fe$_4$ chain (\autoref{fig3}a).

In order to stabilize the pure spin-spiral state we have built a $3a$-chain with one end RKKY-coupled to a ferromagnetic Co stripe (\autoref{fig4}). The Co stripe has a remanent out-of-plane magnetization, which can be used to stabilize the magnetization of atoms that are adsorbed on the Pt(111) at several nanometers lateral distance\cite{Meier2008}. The first atom of the chain was put at a lateral distance of $d_{\rm Co-Fe}\approx \SI{1.5}{\nano\meter}$ to the stripe, where the interaction showed strongest antiferromagnetic coupling\cite{Meier2008}. Starting with this atom, a Fe$_{16}$ chain with $d_{\rm NN}=3a$ aligned almost perpendicular to the rim of the Co stripe has been assembled (\autoref{fig4}a). Since the RKKY-coupling to the stripe decays rather slowly with lateral distance\cite{Meier2008}, we expect possible residual couplings for, say, the $2^{\rm nd}$ and $3^{\rm rd}$ chain atoms, but not for all others, such that an undisturbed spin-spiral can form. The SPSTM image of the chain (\autoref{fig4}b, top panel) taken with a tip with sensitivity to the out-of-plane component of the magnetization (Supplementary Note 5 and Figure 12) reveals an alternating height of neighboring atoms in some areas (cf. atoms 2 to 4, 7 to 9 and 11 to 14) and almost vanishing contrast in between (cf. atoms 5 to 6 and 9 to 11). To prove the magnetic origin of these height differences, the out-of-plane magnetization of the Co stripe was reversed by ramping the field $B_z$ to $\SI{-1.7}{\tesla}$ and back to zero, while the tip was retracted. Due to the RKKY-stabilization, this procedure reverses the magnetization of atom 1 in the chain, and thereby supposedly switches the chain from a state \stateblue{\textbf{0}} to a state \statered{\textbf{1}} with a reversed magnetization of all chain atoms. Please note that the magnetic state of the tip thereby remained unchanged (Supplementary Note 5 and Figure 12). Indeed, in the subsequently taken SPSTM image of state \statered{\textbf{1}} (\autoref{fig4}b, bottom panel) the alternating height contrast is reversed with respect to state \stateblue{\textbf{0}}, while the vanishing contrast regions remain unaffected. For a quantitative analysis of the magnetic contrast, we measure the height on top of each atom ($i$) in state \textbf{\statered{1}} ($h_{i\mathrm{,1}}$) and \textbf{\stateblue{0}} ($h_{i\mathrm{,0}}$) (after background subtraction) and calculate the spin-asymmetry $\Delta_i$ by $h_{i\mathrm{,0}} - h_{i\mathrm{,1}}$ divided by the sum $h_{i\mathrm{,0}} + h_{i\mathrm{,1}}$, which is directly proportional to the spin-polarization of the atoms along the magnetic orientation of the tip\cite{Brede2010}. In \autoref{fig4}c, $\Delta_i$ reveals a long-range beating with a wavelength of about 10 atoms superimposed on the short range antiferromagnetic component (see the gray points in \autoref{fig4}c for which the antiferromagnetic component was removed by reversing the sign of all even numbered atoms). This strongly indicates the stabilization of an antiferromagnetic spin-spiral in the Fe$_{16}$ chain.

This finding is further corroborated in \autoref{fig5}b by a comparison of the Fourier transform of the measured spin-asymmetry $\Delta_i$ with the $z$-component of the spin-structure factor $\big<\hat{S}_z(q)\big>$ obtained from a density-matrix renormalization group (DMRG) calculation for an isolated spin $S = \nicefrac{5}{2}$ chain with the first spin pinned along $z$ (see Methods). We used the same NN coupling parameters as for the ED and perturbation theory model calculations, which were extracted from the experimental investigation of pairs\cite{Khajetoorians2016}. The experimental and theoretical curves show a very similar trend with the formation of a clear maximum at the wavevector $q\approx 7\pi/(8d_{\rm NN})$ (see the dashed vertical line). This finally substantiates the formation of a spin-spiral in the bottom-up constructed Fe$_{16}$ chain with a wavelength of $\lambda = 2\pi/q \approx 2.29 d_{\rm NN} = \SI{19}{\angstrom}$, which is shown in real space in \autoref{fig5}a. The commensuration length of this spin-spiral resulting from $\lambda$ is approximately 7 atoms.


We finally discuss the tunability of the spin-spiral period via the exchange constant ratio $r = \left|\DperpNN\right|/\left|J_{\rm NN}\right|$, which can be tailored by the distance dependent RKKY-interaction as shown above. \autoref{fig5}c and d illustrate the DMRG calculation of $\big<\hat{S}_z(q)\big>$  for long chains ($N = 128$) as a function of $r$ while keeping $\DperpNN$ (c) or $J_{\rm NN}$ (d) constant, revealing how $q$ depends on $r$. $q$ approaches $\pi/(2d_{\rm NN})$ for large $r$, corresponding to a spin-spiral with perpendicular orientations of neighboring spins. For $r\approx 1$, which corresponds to the case experimentally realized here, the peak gets broader due to the competition of Heisenberg and DM interactions, and the magnetic anisotropy, but is still well defined, indicating that the wavevector of the antiferromagnetic ($J_{\rm NN}<0$) spin-spiral can in principle be adjusted within the range $\pi/(2d_{\rm NN})<q<\pi/d_{\rm NN}$. For the other experimentally realized $4a$-chain, the DMRG calculated $\big<\hat{S}_z(q)\big>$ (see cyan curve in \autoref{fig5}b) predicts a maximum at $q \approx 104\pi/(128d_{\rm NN})$ corresponding to an antiferromagnetic spin-spiral with a larger wavelength of $\lambda \approx \SI{27}{\angstrom}$ and opposite rotational sense as compared to the $3a$-chain.

Summarizing, we demonstrated that the rotational sense and the period of spin-spirals realized in bottom-up constructed chains of RKKY-coupled transition-metal atoms on a high-$Z$ substrate can be tuned by the NN distance of the chain atoms. Building chains with other experimentally accessible NN distances will enable to further adjust $J_{\rm NN}$ and $\DperpNN$,\cite{Khajetoorians2016} including ferromagnetic Heisenberg interaction. Our DMRG calculations of according chains using the experimentally determined couplings\cite{Khajetoorians2016} predict that $q$ (see the colored arrows in \autoref{fig5}b, Supplementary Figure 14 and Supplementary Table 1) can thereby be varied from $\pi/(4d_{\rm NN})$ to $114\pi/(128d_{\rm NN})$, corresponding to a large variation in the wavelength from $\lambda \approx \SI{17}{\angstrom}$ to $\SI{133}{\angstrom}$. This tunability of non-collinear spin states could be very important for the manipulation of Majorana end modes of such chains assembled on a superconducting substrate\cite{Nadj-Perge2013, Pientka2013}.

\begin{methods}

\subsection*{Experimental procedures}
All experiments were performed in a home-built STM facility\cite{Wiebe2004} at a temperature of $\SI{0.3}{\kelvin}$. A magnetic field of up to $B_z=\SI{12}{\tesla}$ was applied perpendicular to the sample surface. The Pt(111) sample was cleaned \textit{in-situ} by subsequent sputter-and-anneal cycles as well as O$_2$ annealing and high temperature flashes\cite{Khajetoorians2016, Khajetoorians2013}. Finally the sample was flashed to $\SI{1000}{\celsius}$ for $\SI{1}{\minute}$. In order to prepare the Co stripes, a fraction of a monolayer Co was deposited from an electron-beam heated rod during cooling down the sample to room temperature after the final flash, resulting in the decoration of the Pt step edges and terraces with one atomic layer high Co stripes and islands, respectively\cite{Meier2008}. For the deposition of Fe atoms, about \SI{1}{\percent} of a monolayer Fe was evaporated onto the cold sample kept at $T\lesssim\SI{10}{\kelvin}$.\\

STM topographs were recorded in the constant-current mode of the STM at a stabilization current $I_\mathrm{stab}$ and voltage $V_\mathrm{s}$ applied to the sample. For the assembly of the Fe chains, the Fe atoms were manipulated using lateral tip-induced atom manipulation\cite{Hermenau2017} with a current of $I_\mathrm{stab}\approx \SI{40}{\nano\ampere}$ and a voltage of $V_\mathrm{s} = \SI{2}{\milli \volt}$. We used flashed and Cr coated ($\approx \SI{50}{ML}$) tungsten tips which were prepared as described before\cite{Hermenau2017}. In order to achieve magnetic contrast in the SPSTM images or in SP-ISTS with a sensitivity to the out-of-plane component of the sample magnetization (along $z$), the tip was gently dipped into a Co layer and several single Fe atoms were picked up, until a sufficient contrast was observed on the remanently out-of-plane magnetized Co stripes or islands (see Supplementary Figure 12). For non-magnetic ISTS measurements, voltage pulses were applied to the tip and/or it was gently dipped into the Pt substrate until the magnetic contrast vanished and a spectroscopically flat Pt substrate spectrum within the voltage range used for ISTS was obtained\cite{Hermenau2017}. ISTS was performed by adding a modulation voltage $V_\mathrm{mod}$ (rms) of frequency $f_\mathrm{mod} = \SI{4.142}{\kilo\hertz}$ to $V_\mathrm{s}$. After stabilizing the tip at $I_\mathrm{stab}$ and switching the feedback off, the bias $V_\mathrm{s}$ was ramped from the start to the end voltage of the desired spectroscopy range. Simultaneously, the differential conductance $\didu$ was recorded via a lock-in amplifier. In order to eliminate tip related features, all presented spectra of atoms were normalized by dividing by a spectrum taken with the identical tip on the bare Pt substrate.

\subsection*{Perturbation theory model}
The experimental ISTS data was modeled by employing the perturbation theory code (v0.999) by M. Ternes\cite{Ternes2015}. The model is based on the following effective spin Hamiltonian
\begin{align}
	\begin{split}
		\hat{\Hcal} = &\hat{\Hcal}_\text{Zeeman} + \hat{\Hcal}_\text{anisotropy} + \hat{\Hcal}_\text{Heisenberg} + \hat{\Hcal}_\text{DM} \\
				 = &- \sum^N_i g \muB \mathbf{B} \cdot \hat{\mathbf{S}}_i	+ \sum^N_i K \left( \hat{S}_{i,z} \right)^\text{2} \\
				 &- \sum_{i \neq j}^{N} J_{ij} \, \hat{\mathbf{S}}_i \cdot \hat{\mathbf{S}}_j - \sum^N_{i \neq j} \mathbf{D}_{ij} \left( \hat{\mathbf{S}}_\text{i} \times \hat{\mathbf{S}}_\text{j} \right),
	\end{split} \label{eq:H3rdOrder}
\end{align}
which includes the Zeeman energy and the magnetic anisotropy of all atoms in the chain, quantified by the Land{\'e} g-factor $g$, Bohr magneton $\muB$, and axial magnetic anisotropy constant $K$, respectively, as well as the mutual Heisenberg and the Dzyaloshinskii-Moriya interactions between atoms $i$ and $j$ quantified by the Heisenberg exchange constant $J_{ij}$ and DM vector $\mathbf{D}_{ij}$, respectively. Note, that no third order effects had to be considered to nicely reproduce the data. All model parameters which are not specified in this manuscript are given in Ref. \citenum{Khajetoorians2016}.



\subsection*{ED and DMRG calculations}

To solve the model \autoref{eq:H3rdOrder} for the $S = \nicefrac{5}{2}$ chains up to $N = 4$, we used exact diagonalization (ED). For the longer chains ($N = 16$ and $N = 128$) we used our density-matrix renormalization group (DMRG) code\cite{Schollwoeck2011}. Note, that the Hilbert space of 16 atoms becomes prohibitively large since we are dealing with a spin $S=5/2$. The DMRG method exploits the relatively small entanglement between parts of the system to compress the wavefunction into a matrix product state which is controlled by the bond dimension $\chi$ (a measure for the amount of variational parameters), and yields essentially exact results. The summations are restricted to nearest neighbours (NN).

Our algorithm adaptively chooses the bond dimension $\chi$ using the energy variance $\epsilon = \big|\big<H^2\big>-\big<H\big>^2\big|/N$ as a convergence criterion for the ground state (in addition to the ground-state energy), where $\epsilon<10^{-6}$ can be typically achieved with $\chi \sim 50-100$ for a system of $N=128$ sites. Note, that a well-defined peak in $\big<\hat{S}_z(q)\big>$ is already obtained for $N = 16$, which does not change significantly for $N = 128$ sites (cf. Supplementary Figure 13 in Note 6).

Note, that the DM term in \autoref{eq:H3rdOrder} breaks the SU(2) symmetry of the Heisenberg model, such that all the components $\big<\hat{S}_{\parallel,i}\big>$, $\big<\hat{S}_{\perp,i}\big>$ and $\big<\hat{S}_{z,i}\big>$ are in general nonvanishing and the Hamiltonian is complex. To determine the wavevector $q$ of the spiral in a chain of length $N$, we calculate $\big|\big<\hat{S_{\alpha}}\big>\left(q\right)\big| = 1/N \big|\sum_{n=0}^{N-1} \big<\hat{S}_{\alpha,n}\big> ~ e^{iqn}\big|$ with $\alpha=\parallel,\perp,z$ and $q$ taking the values $q_n=2\pi n/N$, where $n=0,1,\ldots,N-1$. For $\Dpar=0$ and $D_z=0$ the Hamiltonian becomes real, which speeds up the calculations considerably. Furthermore, the ground state can be chosen real as well, meaning that the perpendicular component vanishes: $\big<\hat{S}_{\perp, i}\big>=\text{Im}\big<\hat{S}^+_i\big>=0$.

\end{methods}

\newpage
\noindent \textbf{References}


\begin{addendum}
 \item M. S., A. A. K, J. W., M. P. and R. W. acknowledge funding from the SFB668 and the GrK1286 of the DFG. J. H. and A. A. K. also acknowledge funding from the Emmy Noether Program of the DFG (KH324/1-1). R. R. acknowledges funding from the SFB925. Furthermore, parts of the analysis in this manuscript were done with WSxM\cite{Horcas2007} and several plots were made using the ColorBrewer colormaps\cite{ColorBrewer}.

 \item[Author contributions] J. W., M. S., and A. A. K. designed the experiment. M. S., K. T. T., and J. H. carried out the measurements. M. S. and K. T. T. did the analysis of the experimental data. R. R. performed the DMRG calculations. M. S. and R. R. prepared the figures. M. S., J. W. and R. R. wrote the manuscript. All authors contributed to the discussion and interpretation of the results as well as the discussion of the manuscript.

 \item[Competing Interests] The authors declare that they have no
competing financial interests.

 \item[Correspondence] Correspondence and requests for materials
should be addressed to M. Steinbrecher~(email: manuel.steinbrecher@physik.uni-hamburg.de), J. Wiebe~(email: jwiebe@physnet.uni-hamburg.de) or R. Rausch~(email: rrausch@physnet.uni-hamburg.de).
\end{addendum}

\newpage

\newpage

\begin{figure}[H]
	\includegraphics[width = \columnwidth]{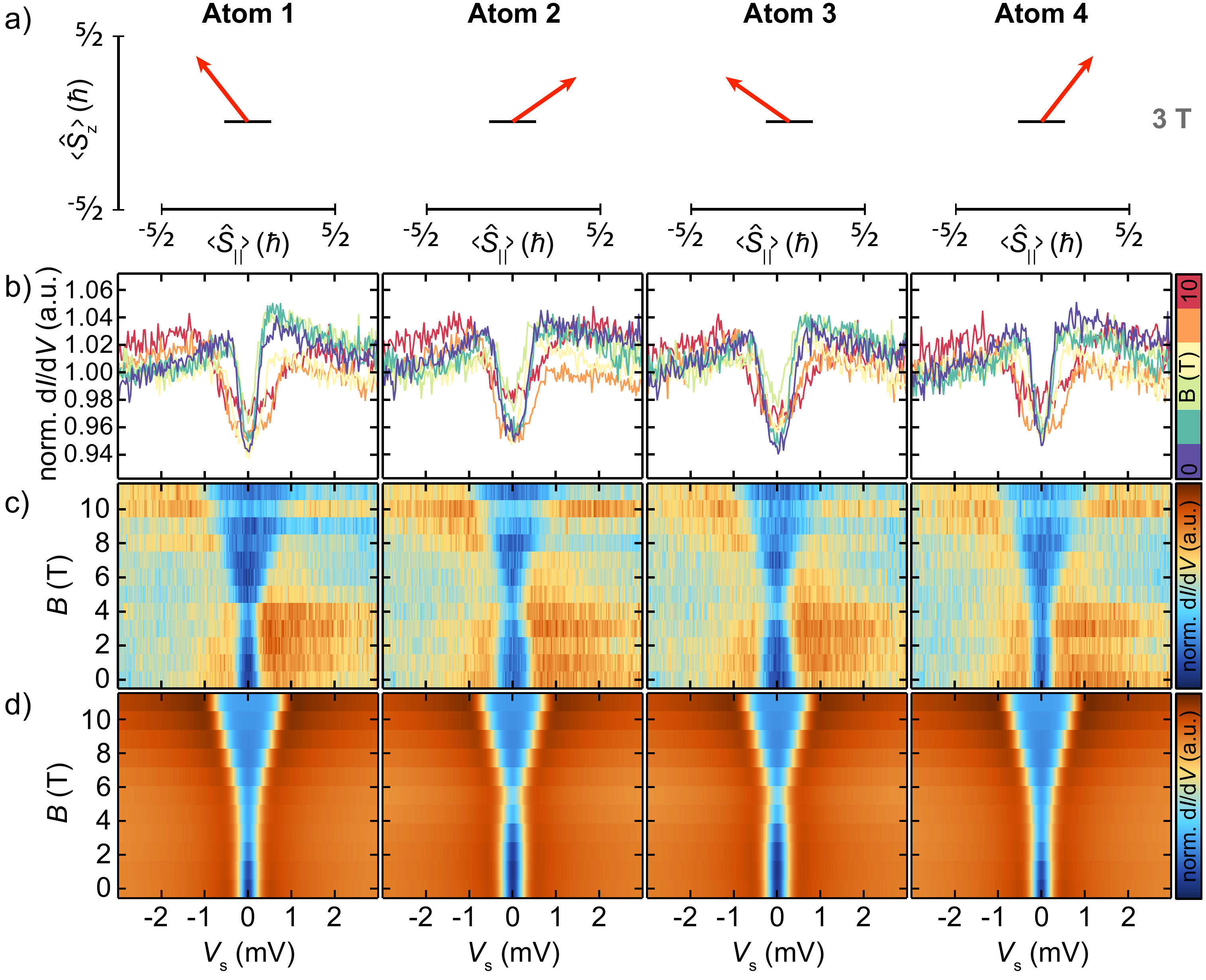} 
 	\caption{\label{fig2} \textbf{Non-collinear spin state in the Fe$_4$ chain with $d_{\rm NN}=4a$.}\\\textbf{a}, Calculated spin-expectation values (\Spar, \Sper, \Sz), see arrows (\Sper$=0$ by symmetry), for each atom in the Fe$_4$ chain at $B_z = \SI{3}{\tesla}$. The results are obtained by exact diagonalization (ED) of a spin model with parameters $J_{\rm NN} = \SI{-25}{\micro\electronvolt}$, $\DperpNN = +\SI{30}{\micro\electronvolt}$, $S=5/2$, $g = 2.0$, and $K = \SI{80}{\micro\electronvolt}$. \textbf{b}, \textbf{c}, ISTS spectra (\textbf{b}) and corresponding colorplots (\textbf{c}) measured on all four atoms in the Fe$_4$ chain for different magnetic fields $B_z$ up to $B_z = \SI{11}{\tesla}$. ($V_\mathrm{stab} = \SI{-6}{\milli \volt}$, $I_{\mathrm{stab}} = \SI{3}{\nano\ampere}$, $V_\mathrm{mod} = \SI{40}{\micro \volt}$). \textbf{d}, Calculated ISTS from the perturbation theory model using the same parameters as in (\textbf{a}).}
 \end{figure}

\newpage

\begin{figure}[H]
	\includegraphics[width = \columnwidth]{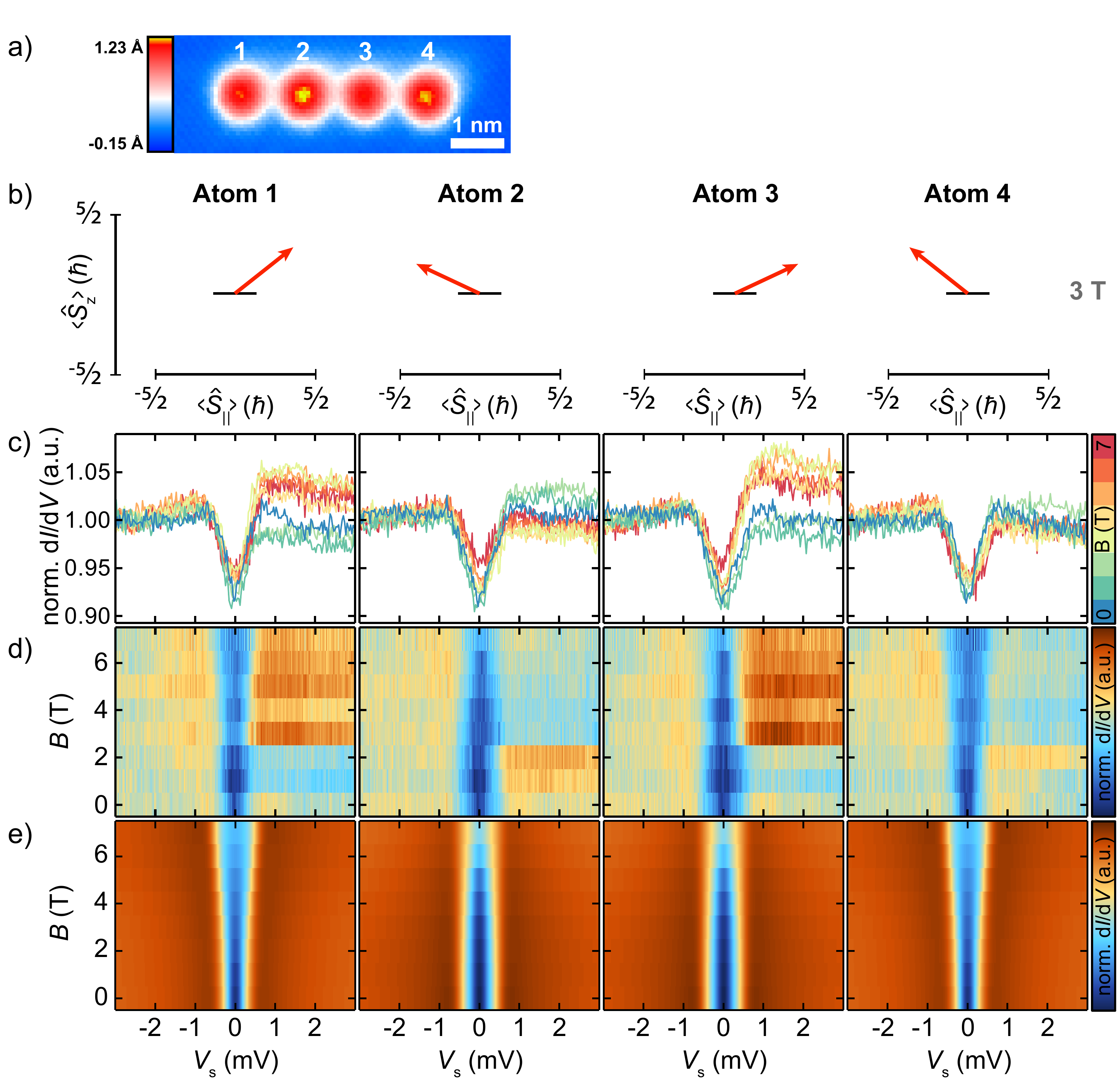} 
 	\caption{\label{fig3}  \textbf{Non-collinear spin state in the Fe$_4$ chain with $d_{\rm NN}=3a$.}\\\textbf{a}, SPSTM topograph of the Fe$_4$ chain measured at $B_z = \SI{3}{\tesla}$ ($V_\mathrm{s} = \SI{-6}{\milli \volt}$, $I_{\mathrm{stab}} = \SI{0.5}{\nano\ampere}$). \textbf{b}, Calculated spin-expectation values at $B_z = \SI{3}{\tesla}$ from ED for each atom in the Fe$_4$ chain, using $J_{\rm NN} = \SI{-60}{\micro\electronvolt}$, $\DperpNN = \SI{-50}{\micro\electronvolt}$, $J_{\rm NNN} = \SI{+15}{\micro\electronvolt}$, $\DperpNNN = \SI{-20}{\micro\electronvolt}$, $S=5/2$, $g = 2.0$, and $K = \SI{80}{\micro\electronvolt}$. \textbf{c}, \textbf{d}, ISTS spectra taken with a magnetic tip (SP-ISTS, \textbf{c}) and corresponding colorplots (\textbf{d}) measured on all four atoms in the Fe$_4$ chain in a magnetic field up to $B_z = \SI{7}{\tesla}$. ($V_\mathrm{stab} = \SI{-6}{\milli \volt}$, $I_{\mathrm{stab}} = \SI{3}{\nano\ampere}$, $V_\mathrm{mod} = \SI{40}{\micro \volt}$). \textbf{e}, Calculated ISTS from the perturbation theory model using the same parameters as in (\textbf{b}).}
 \end{figure}

\newpage

\begin{figure}[H]
\centering
	\includegraphics[width = 0.75\columnwidth]{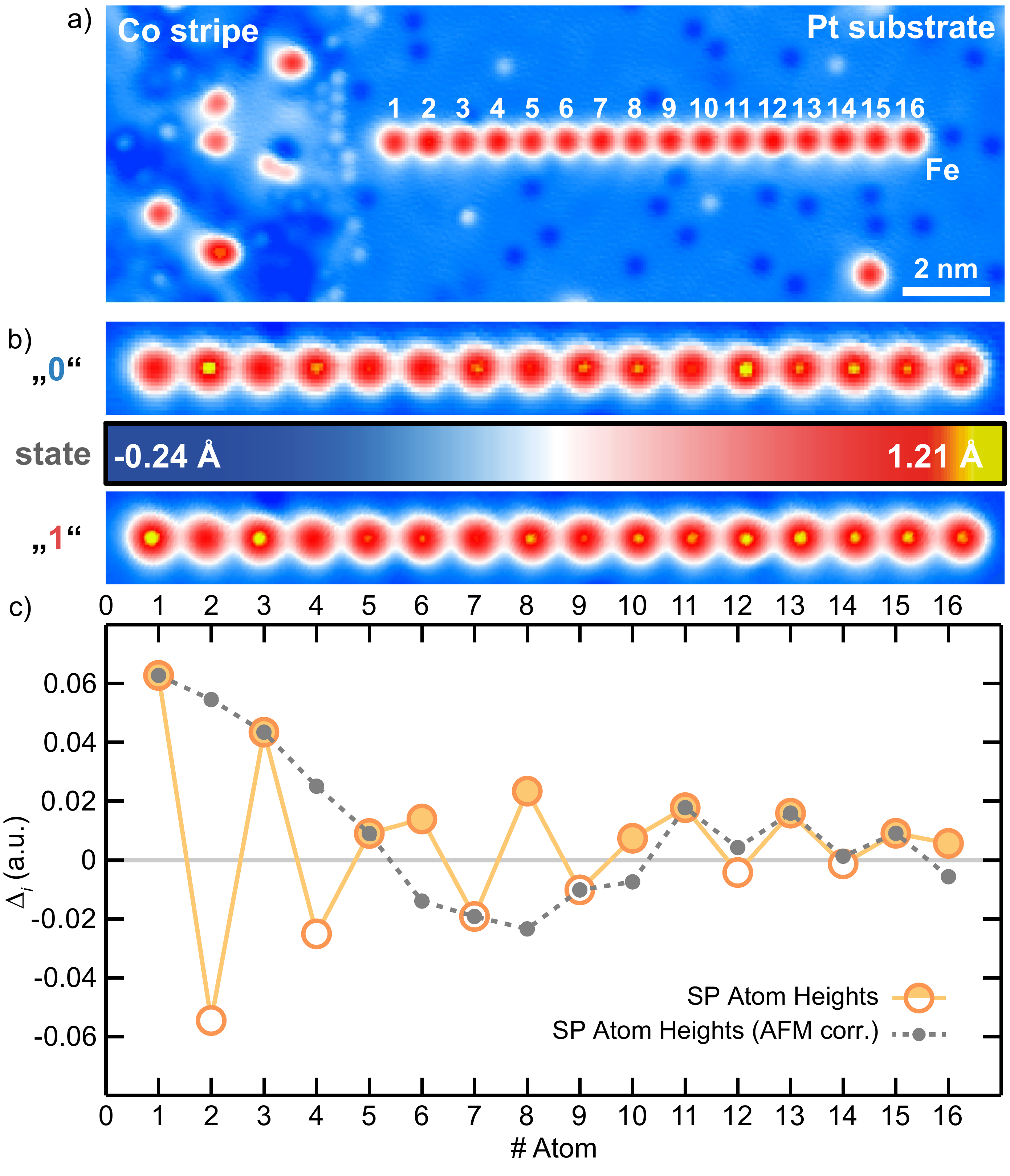} 
 	\caption{\label{fig4}\textbf{Spin-spiral in Fe$_{16}$ chain coupled to a Co stripe.}\\\textbf{a}, STM topograph of an Fe$_{16}$ chain with $d_{\rm NN}=3a$ RKKY-coupled to a ferromagnetic Co stripe on the left. ($I_{\mathrm{s}} = \SI{0.5}{\nano\ampere}$, $V_\mathrm{s} = \SI{6}{\milli \volt}$). \textbf{b}, SPSTM images of the two states ''\stateblue{\textbf{0}}'' and ''\statered{\textbf{1}}'' of the Fe$_{16}$ chain stabilized by the two magnetization states of the Co stripe taken with out-of-plane sensitive tip ($B_z = \SI{0}{\tesla}$, $I_{\mathrm{s}} = \SI{0.5}{\nano\ampere}$, $V_\mathrm{s} = \SI{6}{\milli \volt}$). The magnetic state of the Co stripe was reversed between top and bottom image. \textbf{c}, Spin-asymmetry $\Delta_i$ between states ''\stateblue{\textbf{1}}'' and ''\statered{\textbf{0}}'' of atom $i$ of the Fe$_{16}$ chain taken from the images in (\textbf{b}) (see text). For the gray points, the short-range antiferromagnetic component has been removed by reversing the sign of all even numbered atoms, such that the long range spin-spiral component is visible.}
 \end{figure}

\newpage

\begin{figure}[H]
\centering
	\includegraphics[width = 0.62\columnwidth]{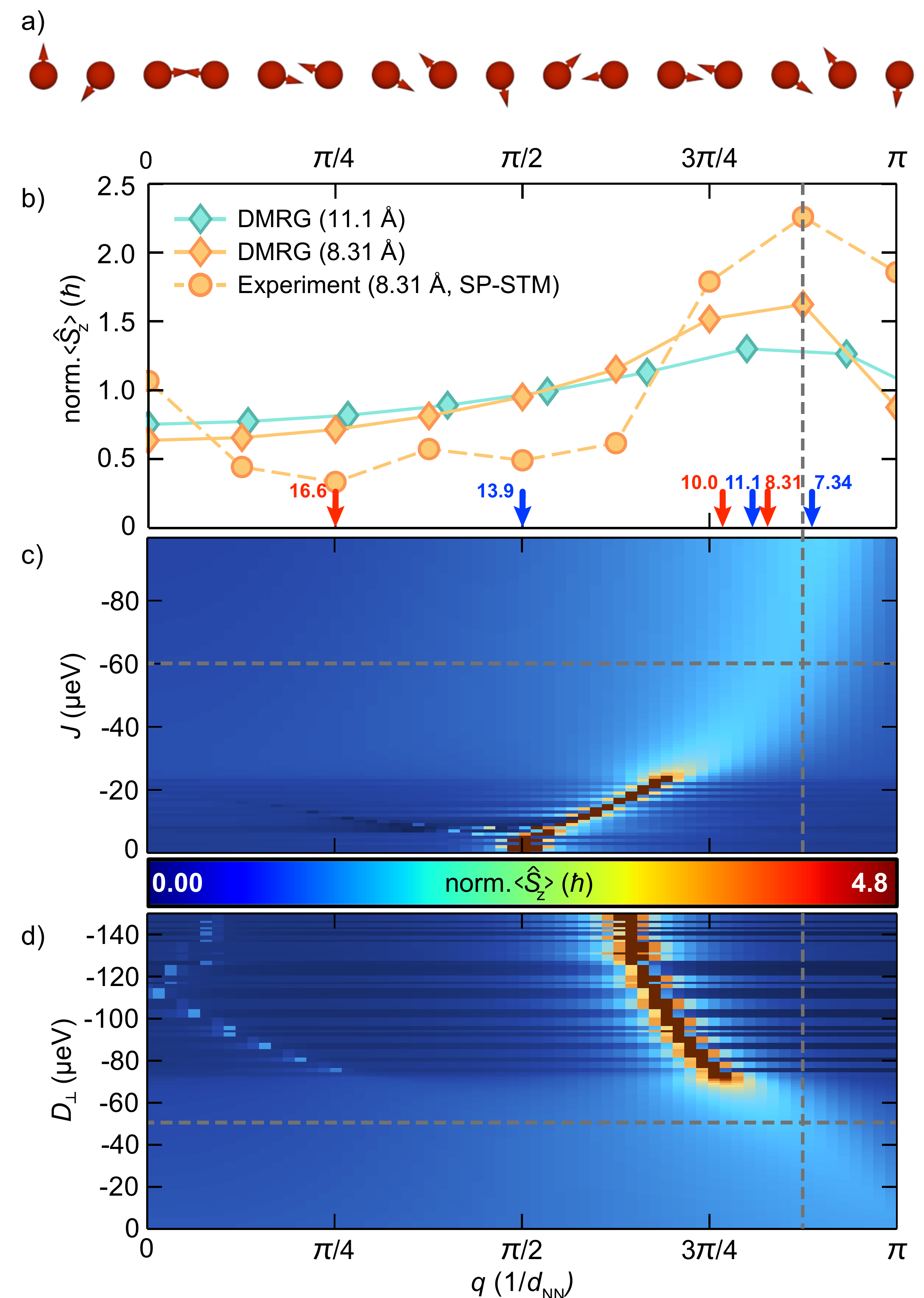} 
 	\caption{\label{fig5}\textbf{Determination of the spin-spiral wavevector from experiment and DMRG.} \\\textbf{a}, Illustration of the spin-spiral realized in the $3a$-chain using the DMRG calculated vectors ($\big<\hat{S}_{\parallel,i}\big>,\big<\hat{S}_{z,i}\big>$) with the length set to unity. \textbf{b}, Comparison of the Fourier transform of the experimental spin-asymmetry (orange dots) taken from Fig.~\ref{fig4} with the spin-structure factor $\big<\hat{S}_z(q)\big>$ obtained from the DMRG calculation of a $d_{\rm NN}=3a$ ($N=16$, orange diamonds) and a $d_{\rm NN}=4a$ ($N=16$, cyan diamonds) chain using the NN interactions experimentally determined from pairs ($d_{\rm NN}=3a$: $J_{\rm NN} = \SI{-60}{\micro\electronvolt}$, $\DperpNN = \SI{-50}{\micro\electronvolt}$; $d_{\rm NN}=4a$: $J_{\rm NN} = \SI{-25}{\micro\electronvolt}$, $\DperpNN = +\SI{30}{\micro\electronvolt}$; $S=5/2$, $g = 2.0$, and $K = \SI{80}{\micro\electronvolt}$). The curves are normalized to the area. The colored arrows indicate the calculated wavevectors $q$ and rotational sense (blue: positive, red: negative) of spin-spirals in chains ($N = 128$) with $d_{\rm NN}$ given in \AA~by the numbers besides the arrows. They were determined from the maxima of the calculated $\big<\hat{S}_z(q)\big>$ using the experimental pair-interactions from Ref. \citenum{Khajetoorians2016} (slightly adjusted $J_{\rm NN}$ for $\dNN = 4a$, see Supplementary Figure 14). \textbf{c}, $\big<\hat{S}_z(q)\big>$ as a function of $J$ with a constant $\Dperp=\SI{-50}{\micro\electronvolt}$ ($N=128$). \textbf{d}, $\big<\hat{S}_z(q)\big>$ as a function of $\Dperp$ with a constant $J=\SI{-60}{\micro\electronvolt}$ ($N=128$). The experimental parameters of the $3a$-chain are indicated by the dashed gray line.}
 \end{figure}

\end{document}